\documentclass[prl,twocolumn,superscriptaddress]{revtex4-1}
\usepackage{amsmath}
\usepackage{amssymb}
\usepackage{color}
\usepackage{graphicx}% Include figure files
\usepackage[colorlinks,bookmarks=false,citecolor=blue,linkcolor=red,urlcolor=blue]{hyperref}

\begin{document}
\title{Non-commutative Dynamic Approaches to the Kibble-Zurek Scaling Limit with an Initial Gapless Order} 
\author{Zhe Wang}
\affiliation{Department of Physics, School of Science and Research Center for Industries of the Future, Westlake University, Hangzhou 310030,  China}
\affiliation{Institute of Natural Sciences, Westlake Institute for Advanced Study, Hangzhou 310024, China}
\author{Chengxiang Ding}
\affiliation{School of Science and Engineering of Mathematics and Physics, Anhui University of Technology,
Maanshan, Anhui 243002, China}
\author{Dongxu Liu}
\affiliation{Department of Physics, School of Science and Research Center for Industries of the Future, Westlake University, Hangzhou 310030,  China}
\affiliation{Institute of Natural Sciences, Westlake Institute for Advanced Study, Hangzhou 310024, China}
\author{Fuxiang Li}
\affiliation{School of Physics and Electronics, Hunan University, Changsha 410082, China}
\author{Zheng Yan}
\email{zhengyan@westlake.edu.cn}
\affiliation{Department of Physics, School of Science and Research Center for Industries of the Future, Westlake University, Hangzhou 310030,  China}
\affiliation{Institute of Natural Sciences, Westlake Institute for Advanced Study, Hangzhou 310024, China}
\author{Shuai Yin}
\email{yinsh6@mail.sysu.edu.cn}
\affiliation{School
of Physics, Sun Yat-sen University, Guangzhou 510275, China}
\affiliation{Guangdong Provincial Key Laboratory of Magnetoelectric Physics and Devices, Sun Yat-sen University, Guangzhou 510275, China}

\begin{abstract}
Nonequilibrium many-body physics is one of the core problems in modern physics, while the dynamical scaling from a gapless phase to the critical point is a most important challenge with very few knowledge so far. In the driven dynamics with a tuning rate $R$ across the quantum critical point (QCP) of a system with size $L$, the finite-time scaling shows that the square of the order parameter $m^2$ obeys a simple scaling relation $m^2\propto R^{2\beta/\nu r}$ in the Kibble-Zurek (KZ) scaling limit with $RL^r\gg1$. Here, by studying the driven critical dynamics from a gapless ordered phase in the bilayer Heisenberg model, we unveil that the approaches to the scaling region dominated by the KZ scaling limit with $RL^r\gg1$ are {\it non-commutative}: this scaling region is inaccessible for large $R$ and finite medium $L$, while merely accessible for large $L$ and moderately finite $R$. We attribute this to the memory effect induced by the finite-size correction in the gapless ordered phase. This non-commutative property makes $m^2$ still strongly depends on the system size and deviates from $m^2\propto R^{2\beta/\nu r}$ even for large $R$. We further show that a similar correction applies to the imaginary-time relaxation dynamics. Our results establish an essential extension of nonequilibrium scaling theory with a gapless ordered initial state.
\end{abstract}

%Usually, simple scaling relations emerge naturally from the complete scaling forms in specific scaling limits close to the quantum critical point (QCP). 

%A scaling relation for large $R$ and finite $L$ is then derived as \(m^2 = m_1 R^{2\beta/\nu r} + m_2 L^{-1} R^{2\beta/\nu r - 1/r}\). Although it bears resemblance to the usual finite-size expansion near equilibrium, they differ profoundly in their physical essence since the second term in the present situation comes from the gapless initial state.

\date{\today}
\maketitle

%How nonequilibrium scaling manifests when starting from a gapless ordered phase with spontaneous continuous symmetry breaking—a ubiquitous scenario in quantum magnetism—remains an open question.

%observe a clear breakdown of the conventional KZ power-law scaling for the order parameter, signaling hybridization of Goldstone modes from the initial phase with the critical fluctuations. While a simple modification of the dynamic exponent initially appears plausible, a full FTS analysis reveals that the process is governed by the original critical exponents of the (2+1)D O(3) universality class.

\textit{\color{blue} Introduction.---} Fathoming the nonequlibrium universal properties of quantum critical points (QCPs) constitutes one of the central research topics in modern physics~\cite{RevModPhys.49.435,Dziarmaga2010Dynamics,Polkovnikov2011Colloquium}. Generally, dynamics of a physical quantitie near a QCP obeys power-law scaling relations, which correspond to specific scaling limits from the full scaling forms encompassing all relevant physical quantities under renormalization group transformation. For instance, for driven critical dynamics across a QCP with a finite rate $R$, the Kibble-Zurek (KZ) mechanism posits a scaling relation between the density of the topological defects $n$ and $R$~\cite{Kibble1976Topology,Zurek1985Cosmological}. The finite-time scaling (FTS) then constructs the scaling forms to the full critical region for general macroscopic quantities and different kinds of driving forces~\cite{Zhong2005Dynamic,PhysRevE.73.047102,Gong2010Finite,Yin2014prb}. Both KZ mechanism and FTS have attracted extensive theoretical and experimental attentions~\cite{Zurek2005Dynamics,Dziarmaga2005Dynamics,Polkovnikov2005Universal,Du2023Kibble,Ko2019Kibble,Maegochi2022Kibble,Keesling2019Quantum,Ebadi2021Quantum,Qiu2020Observation,Ebadi2022Quantum,Sunami2023Universal,Li2023Probing,Zhong2005Dynamic,PhysRevE.73.047102,Deng2008Dynamical,Chandran2012Kibble,Logan2016Universal,Kolodrubetz2012Nonequilibrium,Gong2010Finite,Liu2014Dynamic,Huang2014Kibble,Yin2014prb,Liu2015Quantum,Feng2016Theory,Li2025prl,zeng2025susy,Shu2025Deconfined} and exhibited exceptional utility in the rapidly advancing field of programmable quantum devices~\cite{King2023Quantum,Garcia2024Resolving,Dupont2022Quantum,wang2025tricritical,Wang2025tricritical1}.

%In equilibrium, physical quantities near a QCP obey simple power-law scaling relations, which can be derived in specific scaling limits from the full scaling forms encompassing all relevant physical quantities under renormalization group transformation. It was shown that scaling behaviors emerge not only in equilibrium but also in nonequilibrium systems~\cite{RevModPhys.49.435,Dziarmaga2010Dynamics,Polkovnikov2011Colloquium}. In contrast to the well-established equilibrium theory, however, the theoretical framework for nonequilibrium dynamics is still undergoing rapid development. Extending a systematic scaling theory to characterize nonequilibrium dynamics represents a cutting-edge research challenge and has garnered extensive research attention to date~\cite{RevModPhys.49.435,Dziarmaga2010Dynamics,Polkovnikov2011Colloquium}.

Both the original KZ mechanism and the FTS are based on the adiabatic-impulse scenario, which typically requires a gapped initial state~\cite{Kibble1976Topology,Zurek1985Cosmological,Zhong2005Dynamic,Gong2010Finite,Yin2014prb,Zurek2005Dynamics,Dziarmaga2005Dynamics,Polkovnikov2005Universal}. In particular, starting from a gapped ordered state, the FTS form of the square of the order parameter $m^2$ at the QCP is~\cite{Zhong2005Dynamic,Huang2014Kibble,Feng2016Theory}
\begin{equation}
m^2(R, L) = R^{2\beta/\nu r} f(L^{-1} R^{-1/r}),
\label{mscaling1}
\end{equation} 
in which $\beta$ and $\nu$ are usual critical exponents, $r=z+1/\nu$ denotes the critical dimensionality of $R$ with $z$ being the dynamic exponent, $L$ is the linear size of the system, and $f$ is the scaling function. 

Different scaling limits yield distinct power-law scaling relations, which correspond to different expansions of the underlying FTS form. For instance, when $RL^{r}\ll 1$, the usual finite-size scaling $m^2\propto L^{-2\beta/\nu}$ is recovered from Eq.~(\ref{mscaling1}) as $f(x)\propto x^{2\beta/\nu }$; in contrast, in the opposite limit with $RL^{r}\gg1$~\cite{Zhong2005Dynamic,Huang2014Kibble,Feng2016Theory} 
\begin{equation}
m^2 \propto R^{2\beta/\nu r}.
\label{msrelation}
\end{equation}
Note that the finite-size effects disappear in Eq.~(\ref{msrelation}). By analogy, the original KZM, $n\propto R^{d/r}$~\cite{Kibble1976Topology,Zurek1985Cosmological} ($d$ is the spatial dimension), can also be derived from its scaling when $RL^{r}\gg1$~\cite{Deng2008Dynamical}. Accordingly, this scaling limit will be termed the KZ scaling limit in the following.

Superficially, the same KZ scaling limit $RL^r\gg1$ can be approached in several ways: (i) fixed medium $L$ with large $R$; (ii) fixed medium $R$ with large $L$; and (iii) trivially, simultaneous increase of both quantities. For a gapped initial state, a {\it commutative} property emerges: different approaches to reaching the dominant region of the scaling limit yield identical scaling relations. This {\it commutative} feature offers an efficient route to determine critical properties in finite‑size systems~\cite{Zhong2005Dynamic,Gong2010Finite,Yin2014prb,Huang2014Kibble,Liu2014Dynamic,Liu2015Quantum,Feng2016Theory,Shu2025Deconfined}.

Does this commutative property hold for all systems? Recently, driven critical dynamics from gapless initial states, which breaks the adiabatic-impulse scenario, has garnered considerable attention~\cite{Polkovnikov2008Breakdown,Divakaran2008Quenching,Suzuki2015Universal,Deng2008Dynamical,Zeng2025Finite}. In particular, a universal criterion has been proposed which stipulates that the FTS form, such as Eq.~(\ref{mscaling1}), and the KZ mechanism remain valid if the dynamic exponent $z_0$ of the initial gapless phase satisfies $z_0<r$~\cite{Zeng2025Finite}. This criterion is applicable not only to the Dirac criticality~\cite{Zeng2025Finite} but also accounts for a host of prior findings~\cite{Divakaran2008Quenching,Suzuki2015Universal,Deng2008Dynamical}. Yet key questions emerge: does the validity of the scaling form necessarily entail that of the scaling relation in a given scaling limit? Alternatively, do distinct pathways to approaching the scaling limit yield the same scaling relation?

%Or, particularly, does this imply that the critical dynamics of the order parameter starting from the gapless ordered state can always be described by Eq.~(\ref{msrelation})? 

\begin{figure}[!tb]
\centering
\includegraphics[width=\columnwidth]{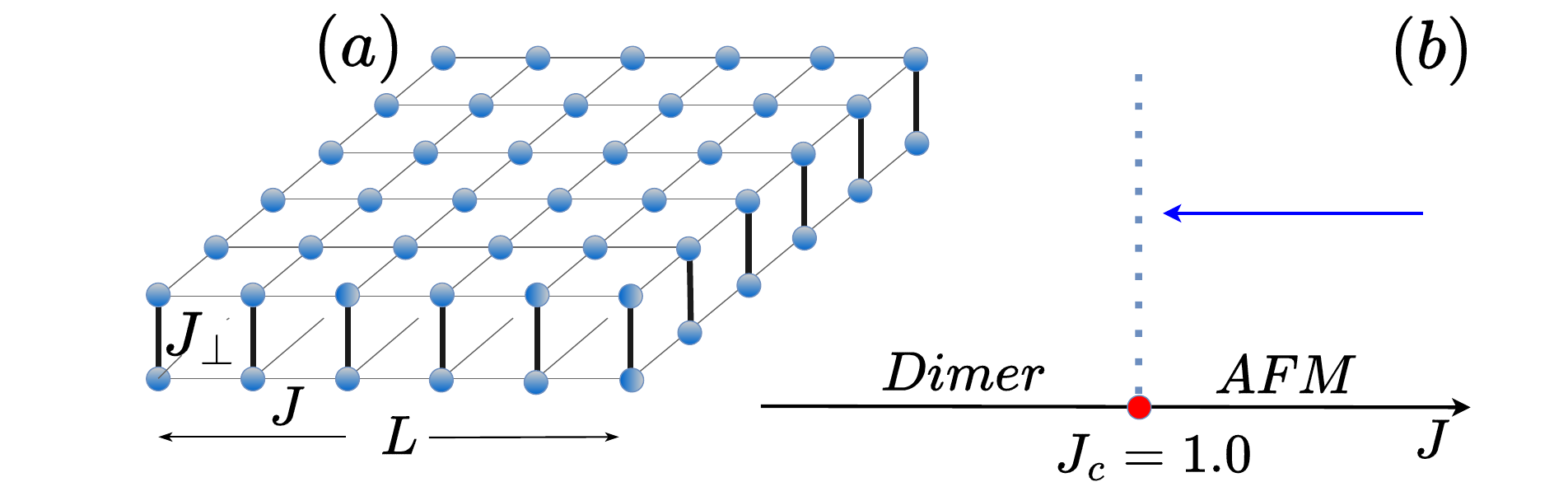}
\caption{Bilayer Heisenberg model. (a) There are two different couplings: $J$ (intraplane) and $J_{\perp}$ (interplane), as indicated. (b) Fixing \( J_{\perp} = 2.522 \), the system is in the gapped dimer phase when \( J < 1.0 \), and in the gapless antiferromagnetic (AFM) phase when \( J> 1.0 \). The blue arrow represents the driving starting from the AFM state.}
\label{fig:model}
\end{figure}

To address these questions, we investigate the driven critical dynamics of the bilayer Heisenberg model, a paradigmatic system hosting a QCP between a gapped dimer phase and a gapless N\'eel phase~\cite{Wang2006High}, as shown in Fig.~\ref{fig:model}. Using large-scale quantum Monte Carlo (QMC) simulations, we drive the system from a gapless antiferromagnetic (AFM) initial state across the QCP. Our findings confirm Eq.~(\ref{mscaling1}) but reveal a clear breakdown of the scaling relation of Eq.~(\ref{msrelation}) in the region with large $R$ and finite $L$. In this way, we show that the approaches to the scaling region dominated by the KZ scaling limit are {\it non-commutative}: this scaling region is inaccessible for large $R$ and finite medium $L$, while only trivially accessible for very large $L$ and finite $R$.

These discoveries lead us to identify a previously overlooked and fundamentally significant issue: distinct pathways to the same scaling limit can give rise to different scaling behaviors, with the origin of this discrepancy lying in the non-negligible size dependence of the gapless initial state~\cite{Neuberger1989Finite,Fisher1989Universality,Hasenfratz1993Finite,Sandvik1997}. This necessitates the inclusion of a size-dependent correction term even for large $R$, leading to the scaling relation of $m^2$ at large $R$ as follows: 
\begin{equation}
m^2 = m_1 R^{2\beta/\nu r} + m_2 L^{-1} R^{2\beta/\nu r - 1/r}.
\label{mscaling2}
\end{equation}
where $m_1$ and $m_2$ are the coefficients for the two terms.

We show that Eq.~(\ref{mscaling2}) can successfully describe the data. Note that the second term here does not stem from the size correction under small $R$, as discussed in Refs.~\cite{Huang2014Kibble,Liu2014Dynamic}, but rather from the memory of the initial-state size dependence at large $R$. We further show that a similar correction enters the critical scaling relations not only in driven dynamics, but also in relaxation dynamics. The {\it non-commutative} approaches to the scaling limit revealed by our study is not only a new finding regarding the fundamental aspects of non-equilibrium scaling theory, but also supplies novel components for corresponding experiments, which all carried out on finite-size systems.

\textit{\color{blue} Models and method.---} Let us focus on the spin-1/2 bilayer Heisenberg model shown in Fig.~\ref{fig:model}~(a). Its Hamiltonian is given by~\cite{Shevchenko2000Double,Wang2006High}
\begin{equation}
H=J\sum_{\alpha=1,2}\sum_{\langle ij\rangle}\vec{S}_{i,\alpha}\vec{S}_{j,\alpha}+J_{\perp}\sum_{i}\vec{S}_{i,1}\cdot\vec{S}_{i,2},
\label{hamil}
\end{equation}
where $\alpha$ is the layer index, and $\langle ij\rangle$ denotes the nearest neighbors in a layer. Both coupling constants are antiferromagnetic ($J,J_{\perp}>0$). For $ J_{\perp}/J \ll 1 $, the ground state is a gapless AFM state with spontaneously $O(3)$ symmetry-breaking, characterized by an AFM order parameter defined as $
m^2 = \frac{1}{L^{2}} \langle (\sum_{i} \epsilon_i S^z_i)^2 \rangle$ 
where $ \epsilon_i = \pm 1 $ depending on whether site $ i $ belongs to the A or B sublattice. In contrast, for $J_{\perp}/J \gg 1$, the ground state approaches a symmetric interlayer-dimerized state with energy gap. The QCP is located at $J_{\perp}/J=2.5220(1)$ and belongs to the (2+1)D O(3) class~\cite{Wang2006High}. 

In this work, we fix $J_{\perp}=2.522$ and employ the nonequilibrium QMC (NEQMC) in the stochastic series expansion (SSE) framework~\cite{sandvik1999stochastic,syljuaasen2002quantum,sandvik2010computational,sandvik2019stochastic,yan2019sweeping,yan2022global} to simulate imaginary-time evolution under time-dependent Hamiltonians. For a time-dependent Hamiltonian $ H(\tau) $, the quantum state evolves according to the Schrödinger equation of imaginary time, such that the evolved state is given by $ |\psi(\tau)\rangle = U(\tau) |\psi(0)\rangle $, where $ |\psi(0)\rangle $ is the initial state at $ \tau = 0 $, and $U(\tau) = \mathcal{T}_\tau \exp\left[ -\int_0^\tau d\tau' \, H(\tau') \right] $ 
is the Euclidean time-evolution operator with $ \mathcal{T}_\tau $ enforcing time ordering~\footnote{Following the standard projector SSE approach, we expand $ U(\tau) $ into a power series of the Hamiltonian terms. The Hamiltonian is decomposed into local bond operators, and by inserting complete sets of basis states along with identity operators associated with time integrals, the expansion is stochastically sampled. The resulting series is truncated at a sufficiently high order to ensure convergence, and importance sampling is performed for the norm $ Z = \langle \psi(\tau) | \psi(\tau) \rangle $ in a chosen basis—typically the $ S^z $-diagonal basis.}. Physical observables are measured at the midpoint of the projection process, where  $\langle A(\tau) \rangle = \frac{\langle \psi(\tau) | A | \psi(\tau) \rangle}{Z}$ and $Z = \langle \psi(\tau) | \psi(\tau) \rangle$. Further technical details of the NEQMC implementation can be found in~\cite{Grandi2011Universal,Grandi2011Universal,Grandi2013Microscopic,Liu2013Quasi,Zeng2025Finite,Shu2025Deconfined}. For driven critical dynamics, it was shown that both the real-time and the imaginary-time dynamics share the same scaling theory~\cite{Grandi2011Universal}.

\begin{figure}[htp]
\centering
\includegraphics[width=0.4\textwidth]{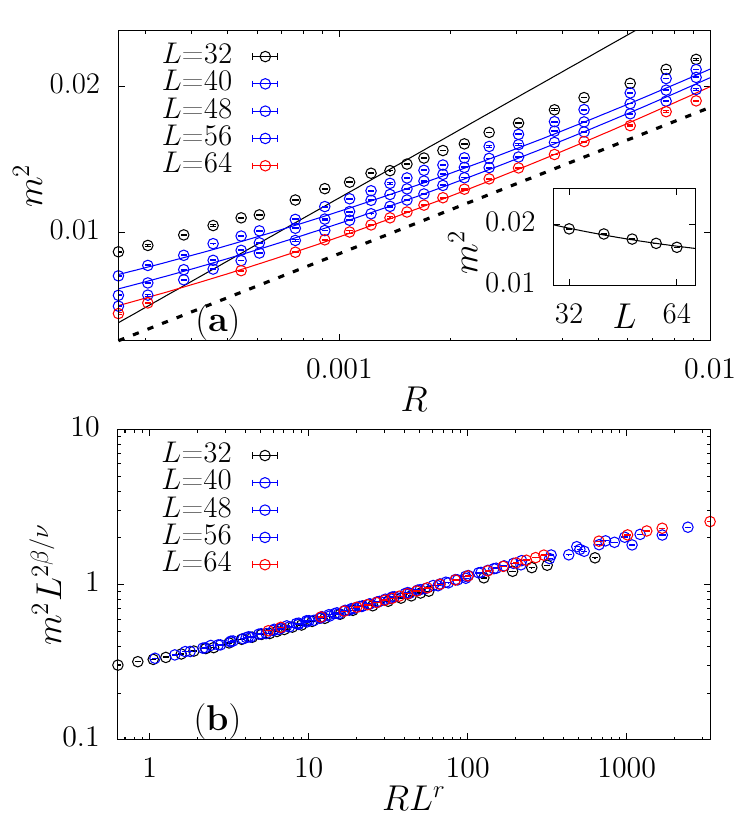}
\caption{Bilayer Heisenberg model: driven dynamics of $m^2 $ versus $R$ for different system size $L$ on log-log plots. From the gapless AFM phase to the QCP using linear  protocol before (a) and after rescaling (b). The black solid line segments in (a) have slopes of $2\beta/(\nu r)$, while the black dashed lines have slopes of $2\beta/\nu r_{02}$. The red and blue curves are fitting lines corresponding to the three largest system sizes, obtained using our newly proposed  scaling relation. The inset in (a) shows $m^2 \propto 0.22\,L^{-1} + 0.012 $ for $R = 0.0045$ (black curve).}
\label{fig:bikz}
\end{figure}

\textit{\color{blue} Driven dynamics.---} We explore the driven quantum dynamics starting from the gapless AFM initial state, where the interaction strength is ramped down in imaginary time according to $ J(\tau) = J_0 - R\tau $, with $ J_0 = 2.522 $. This initial state exhibits AFM long-range order and hosts gapless transverse excitations (Goldstone modes).

We mainly focus on the range near the KZ scaling limit with $RL^r\gg 1$ for medium system sizes. Fig.~\ref{fig:bikz} (a) shows the numerical results of $m^2$ versus $R$ for $L$ from $L=32$ to $64$. It is evident that the data in this region cannot be described by Eq.~(\ref{msrelation}). On one hand, for a single system size, the dependence of $m^2$ on $R$ apparently deviates from the scaling relation of Eq.~(\ref{msrelation}) with $\beta = 0.366(2)$, $ \nu = 0.71(1)$, and $z=1$ belonging to the $(2+1)$D Heisenberg universality class set as input~\cite{Shevchenko2000Double,Wang2006High}, as illustrated in Fig. \ref{fig:bikz} (a) by the black solid curve; on the other hand, these curves display a pronounced size dependence, which contradicts the size-independent behavior predicted by Eq.~(\ref{mscaling1}). Collectively, these findings demand a fresh perspective on the nature of this scaling behavior.

To resolve this puzzle, we first revisit the adiabatic-impulse scenario of the original KZM and FTS, which usually requires a gapped initial state. This initial gap serves two key functions: First, it ensures an initial adiabatic stage with suppressed excitations, confining the dominant non-adiabatic effects to the gap-closing critical region and endowing $R$ with the critical dimensionality of the critical point, thereby characterizing the scaling of driving dynamics. Second, the energy gap of the initial state suppresses the dependence of physical observables on the system size, enabling those at moderate system sizes to approach their thermodynamic-limit values with negligible error. 

%With the gapped initial state, commutative feature for different approaches to the scaling limit $RL^r\gg 1$ appears and Eq.~(\ref{mscaling1}) holds for large $R$ and medium $L$. %responsible for non-adiabatic excitations, 

In contrast to this gapped condition, the primary difference in our present scenario is that the initial state is a gapless ordered state, characterized by two notable features: {\it Feature 1}---gapless excitations induced by any finite driving rate; {\it Feature 2}---inherent finite-size effects in the initial state due to the gapless Goldstone modes. For instance, the order parameter approaches its saturation value in the thermodynamic limit, scaling with the system size via a power law~\cite{Neuberger1989Finite,Fisher1989Universality,Hasenfratz1993Finite,Sandvik1997}. These two factors may explain why the scaling relation in Eq.~(\ref{msrelation}) breaks down, giving rise to the non-commutative ways to the scaling limit. So which of these two potential causes is the key factor? We clarify this point in what follows.

{\it Feature 1} indicates that when the driven dynamics begins from a gapless phase and subsequently crosses the QCP, excitations can be significantly generated not only near the QCP but also throughout the initial gapless regime, in contrast to the conventional scenario where the initial state is gapped~\cite{Dziarmaga2010Dynamics,Polkovnikov2011Colloquium,Polkovnikov2008Breakdown}. Thus, one may expect that excitations in the gapless initial stage may dominate the dynamics, breaking Eq.~(\ref{mscaling1}) and thus violating Eq.~(\ref{msrelation}). 

Actually, for driven dynamics in the gapless phase, the dynamic exponent $z_0$ is the only exponent to characterize the excitation and $R$ has dimension of $1/z_{0}$~\cite{Zeng2025Finite}. For the bilayer Heisenberg model, there are two dynamic exponents $z_0$: the first one is the spin-wave (magnon) excitations with $z_{01}=1$~\cite{Fisher1989Universality,Neuberger1989Finite,Lavalle1998Anomalous}; the other is low-energy excitations analogous to those of a quantum rotor, namely Anderson tower of state with an effective $z_{02}=2$~\cite{Anderson1952An,Weinberg2017Dynamic,mao2025detecting,mao2025sampling}. 

If directly applying Eq.~(\ref{msrelation}), one finds that the evolution of $m^2$ may satisfy $m^2\propto R^{2\beta/\nu r_{02}}$ with $r_{02}=z_{02}+1/\nu$, as shown in Fig.~\ref{fig:bikz}~(a). This result may indicate that $z_{02}$ dominates the critical behavior. However, this fit is highly spurious, because it completely ignores the fact that the curves of $m^2$ versus $R$ have evident size dependence, as shown in the inset of Fig.~\ref{fig:bikz}~(a). Therefore, Eq.~(\ref{msrelation}) cannot be used directly for fitting, since it is only valid under size-independent conditions.

In fact, according to the universal criterion proposed in Ref.~\cite{Zeng2025Finite}, since $z_{01,02}<r$, the driven dynamics should still be controlled by the QCP and described by the usual FTS form of Eq.~(\ref{mscaling1}). To verify it, we rescale $m^2$ and $R$ with $L$ and find that the rescaled curves collapse well according to $ m^2(R, L) = L^{-2\beta/\nu} \mathcal{F}(R L^r) $, which is an incarnation of Eq.~(\ref{mscaling1}), as shown in Fig.~\ref{fig:bikz}~(b)). This FTS form with the scaling correction induced by the irrelevant term is also verified in another model belonging to the same universality~\cite{Liu2025Driven}. Accordingly, these results rule out {\it Feature 1} as a decisive factor responsible for the violation of Eq.~(\ref{msrelation}).

%In the thermodynamic limit ($L \to \infty$), the energies of these rotor states approach that of the ground state according to a power law: $E_{\text{rotor}} - E_{\text{GS}} \sim L^{-z}$, where $z_{02}=d=2$ is the spatial dimension and $L$ denotes the linear system size. These states can be treated as the effective dynamic exponent $z_{02}=2$.

As such, the root cause of the problem lies in the derivation from Eq.~(\ref{mscaling1}) at $g=0$ to Eq.~(\ref{msrelation}), namely the taking of the limit where $RL^r\gg1$. For the gapped ordered state, the scaling region can be reached from several commutative approaches. Here we point out that this commutativity strongly relies on the negligibility of finite-size effects in the initial gap. In contrast, {\it Feature 2} results in the {\it non-commutative} nature of the approaches to the KZ scaling limit. Although for fixed $R$ and very large $L$, Eq.~(\ref{msrelation}) is expected to be recovered, the scaling region controlled by this limit cannot be approached for large $R$ and medium $L$.%stophere

It is worth noting that the AFM N\'eel order parameter strongly depends on the system size due to the presence of Goldstone modes. Specifically, for the AFM phase in our model, this behavior is described by: $m_0^2=m_s^2+aL^{-1}$, where $m_s$ is the saturated order parameter in the thermodynamic limit and $a$ is the coefficient of the size-dependent term~\cite{Neuberger1989Finite,Fisher1989Universality,Hasenfratz1993Finite,Sandvik1997}. For the dynamics of driving the system across the QCP at a relatively large $R$, the universal information of the initial state can be remembered~\cite{Liu2014Dynamic,Huang2014Kibble,Zeng2025Finite}. This can be confirmed in the inset of Fig.~\ref{fig:bikz}~(a), wherein we can clearly observe that $m^2$ has $L^{-1}$-dependence of $L$ for finite $R$. Thus, in this case, even for large $R$, the size-effects cannot be neglected, in sharp contrast to the case with an initial gapped state~\cite{Zhong2005Dynamic,Huang2014Kibble,Feng2016Theory}. 

To be specific, let us consider the scaling property of $m^2$, whose initial value is $m_0^2$ and $m_0^2$ possesses pronounced finite-size
effect. When $z_0<r$, the QCP dominates the driven dynamics and the excitations induced by the external driving in the gapless phase can be neglected~\cite{Zeng2025Finite}. Accordingly, we generalize the scale transformation of $m^2$ at the QCP under driving with finite rate $R$ as follows
\begin{equation}
m^2(R, L, m^2_0) = b^{-2\beta/\nu} m^2\left[ R b^{r},\, L b^{-1},\, U(m^2_0,b) \right],
\label{mstran}
\end{equation}
in which $U(m^2_0,b)$ is the scale transformation of initial $m^2_0$, and $b$ is the rescaling factor. Note that different from usual relevant variable, whose scale transformation follow the simple form such as $g\rightarrow gb^{1/\nu}$, $m^2_0$ transforms as $m^2_0\rightarrow U(m^2_0,b)= m_s^2b^0+aL^{-1}b^1$ since $m^2_0$ contains both the dimensionless saturated term and the size correction term. For the gapped ordered initial state, $U(m^2_0,b)$ recovers a constant~\cite{PhysRevE.73.047102,Zeng2025Finite}.

Then, for large $R$, $R$ dominates and one can obtain the scaling form of $m^2$ as
\begin{equation}
m^2(R, L, m_0) = R^{2\beta/\nu r} f_1\left[ L R^{r},\, U(m_0^2,R^{-1/r}) \right],
\label{mstran1}
\end{equation}
by choosing $b=R^{-1/r}$ in Eq.~(\ref{mstran})~\cite{PhysRevE.73.047102,Zeng2025Finite}. 

Based on the procedure developed in the Ref.~\cite{Zeng2025Finite}, for large $R$, $f_1$ should reflect the universal properties of $U(m_0^2,R^{-1/r})$. Consequently, one obtains a generalized FTS form of $m^2$,
\begin{eqnarray}
m^2(R, g, L) = (R^{2\beta/\nu r}+a'L^{-1}R^{2\beta/\nu r-1/r}) f_2\left( L R^{r} \right),
\label{mstran2}
\end{eqnarray}
%\\ \times 
wherein $a'$ is another nonuniversal constant.

Although Eq.~(\ref{mstran2}) can be converted to Eq.~(\ref{mscaling1}) by absorbing $(1+aL^{-1}R^{-1/r})$ into $f_2$, the non-communicative property for the approach to the scaling limit can be directly seen from Eq.~(\ref{mstran2}). Specifically, $L$ inside $f_2$ represents the finite-size effects in the equilibrium limit, whereas $L$ in the pre-factor leading term before $f_2$ accounts for the influence of the initial state. For large $R$, $f_2$ tends to a constant and only the size contribution within $f_2$ is negligible; while the contribution outside the function is not, giving rise to Eq.~(\ref{mscaling2}) and reflecting the non-commutative property for large $R$ with medium $L$.

%For the gapped ordered initial state, $U(m^2_0,b)$ recovers a............................... constant~\cite{PhysRevE.73.047102,Zeng2025Finite}.

\begin{ruledtabular}
\begin{table}[!h]
\caption{Fitting results for the data in Fig.~\ref{fig:bikz}~(a) with Eq. (\ref{mscaling2}). We fixed known exponents $\beta$, $\nu$ and $z$ in the fit and treated $ m_1 $ and $ m_2 $ as free parameters. }
\begin{tabular}{l c c  }
 	   	   $L$ 	  & $m_1$ 	  & $m_2$ 	 \\
 	   	   	
\hline
	$48$			& 0.12(1)   	&0.258(8)       \\   
	$56$			& 0.117(4)  	&0.27(1)         \\
    $64$			& 0.115(4)   	&0.27(1)       \\
\end{tabular}
\label{ext1}
\end{table}
\end{ruledtabular}

To verify the scaling analyses discussed above, we fit the numerical results for different $L$ in Fig.~\ref{fig:bikz} (a) according to Eq.~(\ref{mscaling2}). The fitted results are shown in Table.~\ref{ext1}. From Table.~\ref{ext1}, one finds that both $m_1$ and $m_2$ have consistent values within the error bar for different $L$, confirming that for large $R$ and medium $L$, $f_2$ tends to a constant at $g=0$. 

It should be noted that driven dynamics starting from the AFM phase has also been studied for Dirac criticality~\cite{Zeng2025Finite} and deconfined quantum criticality~\cite{Shu2025Deconfined}. However, the non-commutative dynamical scaling property is not observed in these systems. This may arise because the coefficient of $L^{-1}$ in the initial state is relatively small in those setups. Nevertheless, such non-commutative behavior is expected to be universal for driven dynamics with an initial state featuring continuous symmetry breaking. In the Supplemental Material, we further demonstrate this behavior in the driven critical dynamics of an additional model.

\textit{\color{blue}Imaginary time relaxation dynamics.---}Besides the driven critical dynamics, the universality of the non-commutative properties in approaching the scaling limit also manifests itself in the imaginary-time relaxation quantum critical dynamics. 

It was shown that universal dynamic scaling behaviors appear in the imaginary-time relaxation process in quantum criticality~\cite{Yin2014}, similar to the short-time dynamics in classical systems~\cite{Janssen1989New,Li1995Dynamic,Zheng1996Generalized,Zheng1998Generalized}. Recently, investigations into imaginary-time relaxation critical dynamics have expanded to a diverse range of quantum many-body models~\cite{Yin2014pre,Zhong2021prb,Yin2022prl,Shu2023prb,Yu2025Nonequilibrium,Yu2026Preempting,zhang2025Magnetic,Shen2025Universal} and programmable quantum devices~\cite{Zhang2024prb}, showcasing its powerful versatility in probing critical properties. For the Heiseberg universality class, it was shown that short-time scaling corrections should be included~\cite{Cai2024prb}.
\begin{figure}[htp]
\centering
\includegraphics[width=0.4\textwidth]{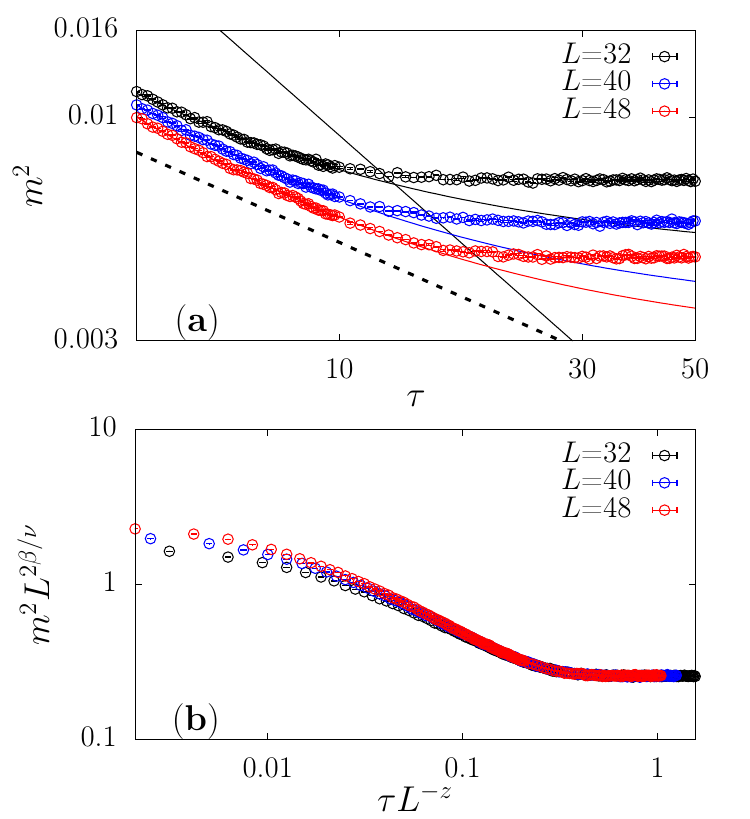}
\caption{Relaxation dynamics of $m^2 $ versus $\tau$ for different system size $L$ on log-log plots.  For a sudden quench from the gapless AFM phase to the QCP before (a) and after rescaling (b), the black solid line in (a) follows the conventional scaling $ m^2 \propto \tau^{2\beta/(z\nu)} $, while the black dashed lines is given by the modified scaling $ m^2 \propto \tau^{2\beta/(z_{\text{02}}\nu)} $.  The red, blue and black curves are fitting lines  obtained using our newly proposed  scaling formula.}
\label{fig:relaxd}
\end{figure}

Here, we prepare the initial state at $ J_0 = 2.522 $ in the gapless AFM phase and then perform a sudden quench to the critical point at $ J = 1.0 $, studying the subsequent imaginary-time relaxation of the order parameter. We find that the short-time scaling relation $ m^2 \propto \tau^{-2\beta/z\nu} $ with a saturated initial order parameter~\cite{Yin2014,Cai2024prb} fails to describe the data (as shown by the black solid line in Fig.~\ref{fig:relaxd}(a)). Although replacing $z$ with $z_{02}$ seems appropriate (black dashed curve), scaling collapse in the relaxation process shown in  Fig.~\ref{fig:relaxd}(b)) demonstrate that the dynamic exponent should still be $z$. 

%Analogous to the driven case, replacing the dynamic critical exponent $ z $ with an effective value of $z_{\text{eff}}=2$ yields a reasonable fit to the data (black dashed line in Fig.~\ref{fig:relaxd}(c)), suggesting an apparent deviation from standard scaling. However, this contradicts the data collapse analysis, which clearly shows that setting $ z = 1 $—the known dynamic exponent at the critical point—leads to excellent scaling collapse across different system sizes ( Fig.~\ref{fig:relaxd}(d)). 

To resolve this discrepancy, and in analogy with our analysis of driven dynamics, a modified scaling relation by incorporating the influence of the finite-size correction in the gapless ordered initial state is then developed: $ m^2 = q_1\tau^{-2\beta/\nu z} + q_2L^{-1}\tau^{-2\beta/\nu z+1/z} $. We find that this scaling relation describes the data very well (see Fig.~\ref{fig:relaxd}(a)). Moreover, the fitted parameters $ q_1 $ and $ q_2 $  remain stable within error bars across different system sizes (see Table~\ref{ext2}), providing strong support for the validity of our scaling ansatz.

\begin{ruledtabular}
\begin{table}[!h]
\caption{Fitting results for the data in Fig.~\ref{fig:relaxd}~(a)  with $ m^2 = q_1\tau^{-2\beta/\nu z} + q_2L^{-1}\tau^{-2\beta/\nu z+1/z} $.	We fixed  known exponents $\beta$, $\nu$ and $z$ in the fit and treated $ q_1 $ and $ q_2 $ as free parameters. }
\begin{tabular}{l c c  }
 	   	   $L$ 	  & $q_1$ 	  & $q_2$ 	 \\
 	   	   	
\hline
	$32$			& 0.025(1)   	&0.181(9)       \\   
	$40$			& 0.0282(3)  	&0.167(4)         \\
    $48$			& 0.0279(4)   	&0.170(3)       \\
\end{tabular}
\label{ext2}
\end{table}
\end{ruledtabular}

\textit{\color{blue} Conclusion and discussion.---} We have uncovered non-commutative approaches to the scaling limit in driven critical dynamics across the quantum critical point from a gapless ordered phase. We find that, while the full finite-time scaling form remains valid, the scaling relation in the Kibble-Zurek scaling limit with $RL^r\gg1 $ breaks down for large $R$ and moderate $L$. We attribute this to the strong system-size dependence of the order parameter in the gapless ordered initial state. We further generalize this non-commutative scaling behavior to imaginary-time relaxation critical dynamics.

Continuous symmetry breaking is a fundamental topic in phase transitions and critical phenomena. Therefore, our scaling formula corrects the previously scaling relation of nonequilibrium driving dynamics at a basic and important level, thereby contributing to a more complete framework of nonequilibrium scaling.

%We have studied non-equilibrium scaling when a system is driven from a gapless ordered phase across a QCP. Using the bilayer Heisenberg model, we observe a breakdown of the conventional Kibble‑Zurek power law for the order parameter—a hallmark of hybridization between Goldstone modes and critical fluctuations. Crucially, the underlying critical exponents remain unchanged, as confirmed by finite-time scaling. Instead, the effect of the gapless initial phase enters as a distinct additive correction in the scaling function, originating from the finite‑size dependence of the order parameter in a symmetry‑broken state. This correction applies both to driven and relaxation dynamics, establishing a necessary refinement of non‑equilibrium scaling theory for gapless ordered phases. Our results provide a key conceptual advance for understanding quantum dynamics in rapidly advancing programmable quantum devices

%Future studies could further explore other quantum critical systems with gapless initial states—such as spin liquids or topological phase transitions—to test the universality of the proposed hybrid scaling theory. Moreover, experimental observation and verification of this hybrid KZM behavior could be realized in quantum simulation platforms such as ultracold atoms and Rydberg atom arrays, by precisely engineering the initial state and quench protocol.

\begin{acknowledgements}
S. Y. is supported by the National Natural Science Foundation of China (Grants No. 12222515 and No. 12075324), Research Center for Magnetoelectric Physics of Guangdong Province (Grant No. 2024B0303390001), the Guangdong Provincial Key Laboratory of Magnetoelectric Physics and Devices (Grant No. 2022B1212010008), and the Science and Technology Projects in Guangzhou City (Grant No. 2025A04J5408).
Z.W. thanks the China Postdoctoral Science Foundation under Grants No.2024M752898. The work is supported by the Scientific Research Project (No.WU2025B011) and the Start-up Funding of Westlake University.
The authors thank the high-performance computing center of Westlake University for providing HPC resources.
\end{acknowledgements}

\bibliography{kz}

\clearpage
\appendix
\setcounter{equation}{0}
\setcounter{figure}{0}
\renewcommand{\theequation}{S\arabic{equation}}
\renewcommand{\thefigure}{S\arabic{figure}}
\setcounter{page}{1}
\begin{widetext}
%\linespread{1.05}
	
\centerline{\bf\Large Supplemental Materials}

\section{Bilayer Heisenberg model} 

\subsection{Driven dynamics from the dimer phase}

\begin{figure*}[htp]
\centering
\includegraphics[width=0.8\textwidth]{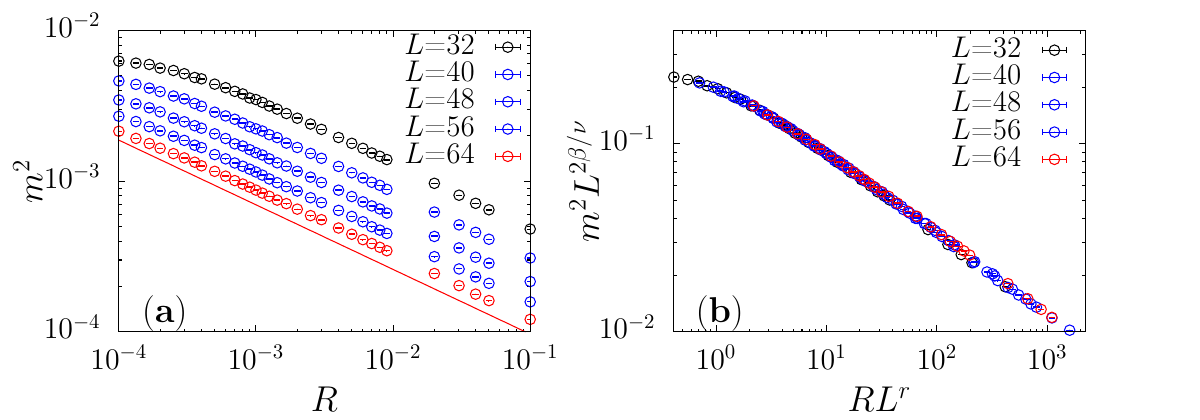}
\caption{ Bilayer Heisenberg model: driven dynamics of $m^2 $ versus $R$ for different system size $L$ on log-log plots. From the dimer phase to the QCP using linear   protocol before (a) and after rescaling (b). The red line segments in (a) have slopes of ($2\beta-d\nu$)/($\nu r$). }
\label{fig:bikzsm}
\end{figure*}

Here, we investigate the driven critical dynamics by tuning the control parameter $ J$ as a function of imaginary time $ \tau $, according to $ J(\tau) = J_0 + R\tau $, starting from a gapped dimer initial state with $ J_0 = 0 $. The distance to the QCP is defined as $ g = J- J_c $, where $ J_c $ is the critical coupling. At the critical point ($ g = 0 $), the order parameter squared $ m^2 $ exhibits distinct scaling behaviors depending on the quench rate $ R $. As shown in Fig.~\ref{fig:bikzsm}~(a), in the fast-quench regime (large $ R $), we observe a power-law decay $ m^2 \propto R^{-q} $ for  a fixed system size $L$, where the exponent $ q$ is derived from the theoretical prediction $q=(2\beta - d\nu)/(\nu r)$~\cite{Liu2014Dynamic,Huang2014Kibble}. Here, $ \beta = 0.366(2) $ and $ \nu = 0.71(1) $ are the critical exponents for the order parameter and correlation length, respectively~\cite{Shevchenko2000Double,Wang2006High}, $ d = 2 $ is the spatial dimension, and $ r = z + 1/\nu $ is the scaling dimension of the quench rate $R $~\cite{Liu2014Dynamic,Huang2014Kibble}. The dynamic critical exponent $ z $ takes the value $ z = 1 $ at (2+1)D O(3) QCP due to emergent Lorentz invariance in the low-energy effective theory~\cite{Shevchenko2000Double,Wang2006High}. This indicates that quenching from the gapped dimer phase to the critical point, the fast-driving regime (large $ R $) is fully described by  the standard FTS relation.

\begin{figure*}[htp]
\centering
\includegraphics[width=0.8\textwidth]{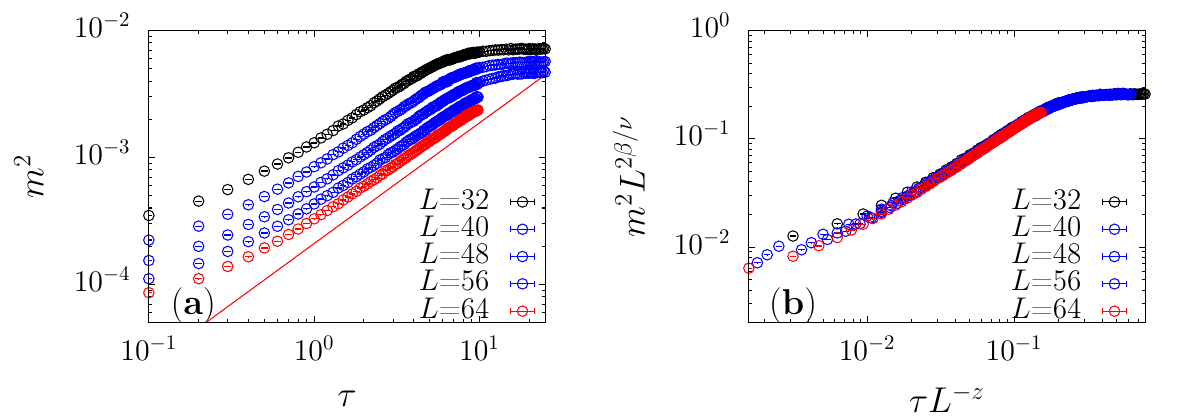}
\caption{Relaxation dynamics of $m^2 $ versus $\tau$ for different system size $L$ on log-log plots.  For a sudden quench from the gapped dimer initial state to the  QCP before (a) and after rescaling (b), the red line in (a) is described by the scaling relation $ m^2 \propto \tau^{(d - 2\beta/\nu)/z} $. }
\label{fig:relaxdsm}
\end{figure*}

Moreover, we find that the entire driving process is indeed well described by the FTS form, expressed as $ m^2(R, L) = L^{-2\beta/\nu} \mathcal{F}(R L^r) $ at $ g = 0 $~\cite{Zhong2005Dynamic,Gong2010Finite,Yin2014prb}. Here, we rescale $ m^2 $ and $ R $ as $ m^2 L^{2\beta/\nu} $ and $ R L^r $, respectively, with the exponents $ \beta = 0.366 $, $ \nu = 0.71$, and $ z = 1 $ fixed as input. The rescaled data for different system sizes collapse onto a single universal curve, as shown in Fig.~\ref{fig:bikzsm}~(b).

\subsection{Imaginary time relaxation dynamics}
We then consider the relaxation dynamics.  We  consider a sudden quench from the gapped dimer phase ($ J_0 = 0.0 $) to the critical point ($ J = 1.0 $). In this case, the data are well described by the  scaling relation $ m^2 \propto \tau^{(d - 2\beta/\nu)/z} $~\cite{Yin2014,Yin2014pre,Zhong2021prb,Yin2022prl,Shu2023prb,Yu2025Nonequilibrium,Yu2026Preempting,zhang2025Magnetic,Shen2025Universal,Zhang2024prb,Cai2024prb} (red solid line in Fig.~\ref{fig:relaxdsm}(a)), and data from all system sizes collapse perfectly when using $ z = 1 $ (Fig.~\ref{fig:relaxdsm}(b)).

\begin{figure}[htb]
\centering
\includegraphics[width=0.6\columnwidth]{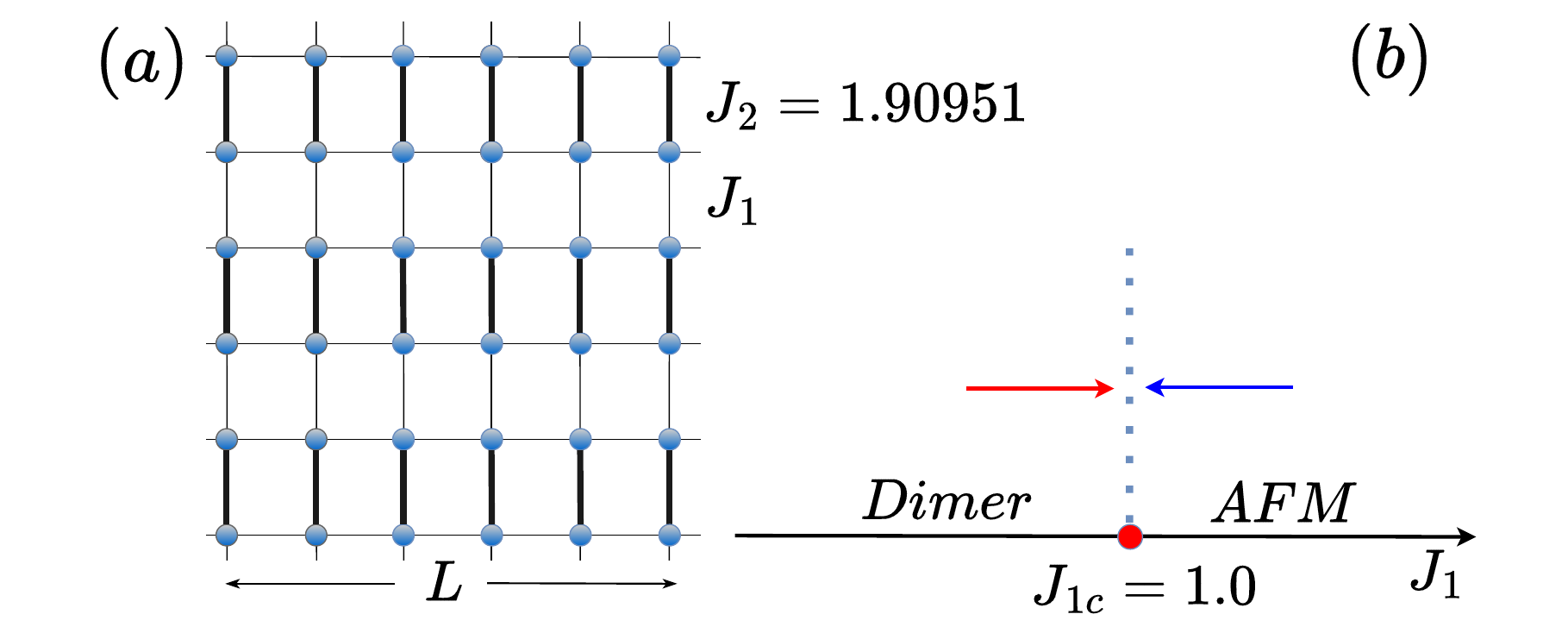}
\caption{Dimerized spin-$1/2$ Heisenberg model on a bipartite 
	square lattice. (a) Strong bonds $J_2>0$ are marked by thick  lines, weak bonds $J_1>0$ by 
	thin  lines. (b) Fixing \( J_{2} = 1.90951 \), the system is in the gapped dimer phase when \( J < 1.0 \), and in the gapless antiferromagnetic (AFM) phase when \( J> 1.0 \). The arrows  represent the driving starting from the AFM state (blue arrow) and dimer state (red arrow), respectively.}
\label{fig:modelsm}
\end{figure}

\section{Dimerized spin-$1/2$ Heisenberg model}

 Let us focus on the dimerized spin-$1/2$ Heisenberg model on a bipartite square lattice, shown in Fig.~\ref{fig:modelsm}~(a). Its Hamiltonian is given by~\cite{Matsumoto2001Ground,Wenzel2008Evidence,Ma2018Anomalous}
\begin{equation}
H=J_{1}\sum_{\langle ij\rangle}S_{i}S_{j}+J_{2}\sum_{\langle ij\rangle'}S_{i}S_{j},
\label{eq:hei}
\end{equation}
in which the summation in the $J_{1}$-term is taken over the thin bonds, while the $J_{2}$-term is over the thick bonds. For $J_{2}/J_{1}\simeq 1$, the ground state is a gapless AFM state, which spontaneously breaks the symmetry from O(3) to O(2). In this phase, the gapless excitations arise due to the presence of Goldstone modes. For the $J_{2}/J_{1}\gg 1$, the ground state approaches a direct product of the spin-singlet states on the thick bonds, which has an energy gap and does not break any symmetries. The QCP separating these two phases is located at $J_{2}/J_1=1.90951$ and belongs to the (2+1)D O(3) class~\cite{Matsumoto2001Ground,Wenzel2008Evidence,Ma2018Anomalous}.

\subsection{Driven dynamics}

Next, similar to the analysis in the main text, we study the driven dynamics and relaxation dynamics of this model. We first focus on the driven dynamics of the system. It is evident that the data (from gapless AFM state to QCP) cannot be described by the black solid line (see Fig.~\ref{fig:j1j2drivengapless}~(a)), whose slope is calculated from the  scaling relation: $ 2\beta / (\nu r) $ with $ r = z + 1/\nu $. We can see that the data are well described by the black dashed line (see Fig.~\ref{fig:j1j2drivengapless}~(a)), whose slope is determined by $ 2\beta / (\nu r_{02}) $. However, we find that the entire driving process is indeed well described by the FTS form (see Fig.~\ref{fig:j1j2drivengapless}~(b)), expressed as $ m^2(R, L) = L^{-2\beta/\nu} \mathcal{F}(R L^r) $, with $ r = 1 + 1/\nu $. For this model, the FTS form with the scaling correction induced by the irrelevant term is also verified~\cite{Liu2025Driven}.

\begin{figure*}[htp]
\centering
\includegraphics[width=0.8\textwidth]{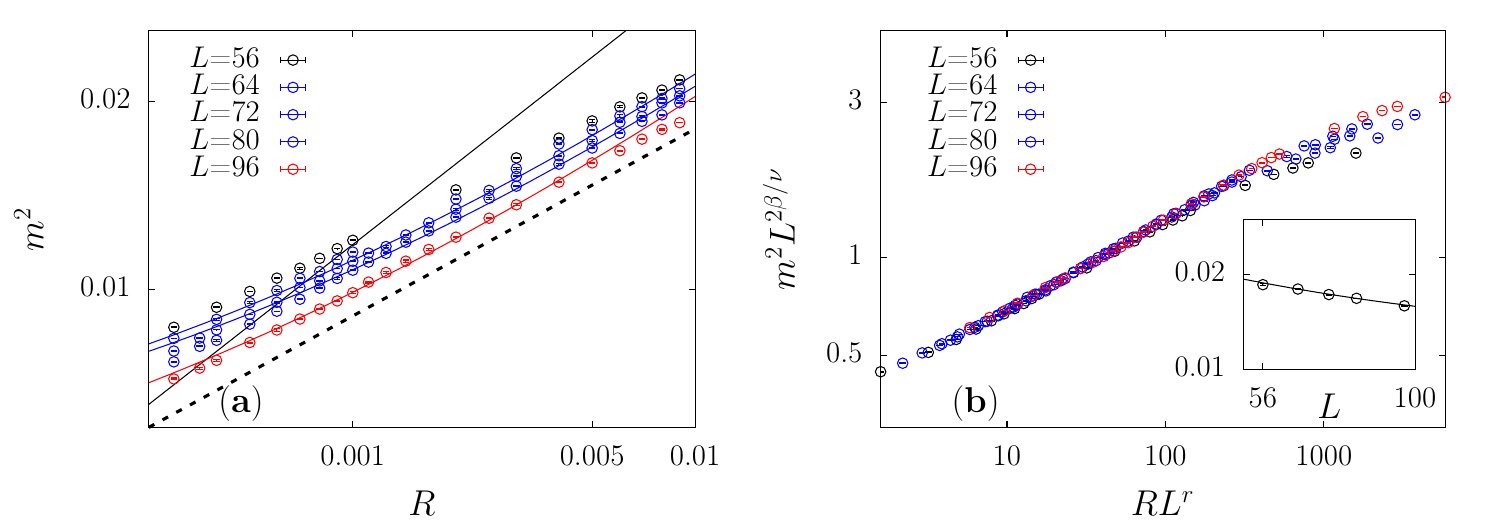}
\caption{ Dimerized  Heisenberg model: driven dynamics of $m^2 $ versus $R$ for different system size $L$ on log-log plots.   From the gapless AFM phase to the QCP using linear  protocol before (a) and after rescaling (b). The black line segments in (a) have slopes of $2\beta/(\nu r)$, while the black dashed lines have slopes of $2\beta/(\nu r_{\text{02}})$. The red and blue curves are fitting lines corresponding to the three largest system sizes, obtained using our newly proposed  scaling relation.}
\label{fig:j1j2drivengapless}
\end{figure*}

\begin{ruledtabular}
\begin{table}[!h]
\caption{Fitting results for the data in Fig.~\ref{fig:j1j2drivengapless}~(a)  with $m^2 = m_1 \, R^{2\beta/(\nu r)}+ m_2 L^{-1} \, R^{2\beta/(\nu r) - 1/r}$.	We fixed  known exponents $\beta$, $\nu$ and $z$ in the fit and treated $ m_1 $ and $ m_2 $ as free parameters. }
\begin{tabular}{l c c  }
 	   	   $L$ 	  & $m_1$ 	  & $m_2$ 	 \\
 	   	   	
\hline
	$72$			& 0.125(3)   	&0.37(1)       \\   
	$80$			& 0.118(4)  	&0.41(2)         \\
    $96$			& 0.119(3)   	&0.40(1)       \\
\end{tabular}
\label{sext1}
\end{table}
\end{ruledtabular}

Next, we re-analyze our data in Fig.~\ref{fig:j1j2drivengapless}~(a) by fitting it to our new scaling relation: $m^2 = m_1 \, R^{2\beta/(\nu r)}+ m_2 L^{-1} \, R^{2\beta/(\nu r) - 1/r}$. We first attempted to fit the data by treating $ m_1 $, $ m_2 $, $ \beta $, $ \nu $, and $ z $ as free parameters, but found the fitting to be unstable. As discussed above (see Fig.~\ref{fig:j1j2drivengapless}~(b)), the $ m^2 $ data across the entire driving process can be successfully collapsed using the universal exponents of the (2+1)D O(3) critical point: $ \beta = 0.366 $, $ \nu = 0.71 $, and $ z = 1 $. Therefore, we fixed these known exponents in the fit and treated only $ m_1 $ and $ m_2 $ as free parameters.  
We find that our proposed FTS scaling relation describes the data very well (see Fig.~\ref{fig:j1j2drivengapless}~(a)). Moreover, the fitted values of $ m_1 $ and $ m_2 $ are stable within error bars across different system sizes(see Table.~\ref{sext1}. This consistency strongly supports the validity of our scaling ansatz.

 If starting from the gapped dimer phase, the results are consistent with those of the bilayer Heisenberg model, both obeying the scaling relation: $m^2 \propto R^{2\beta/\nu r}$ (see Fig.~\ref{fig:j1j2drivengap}~(a)).

 \begin{figure}[htp]
\centering
\includegraphics[width=0.8\textwidth]{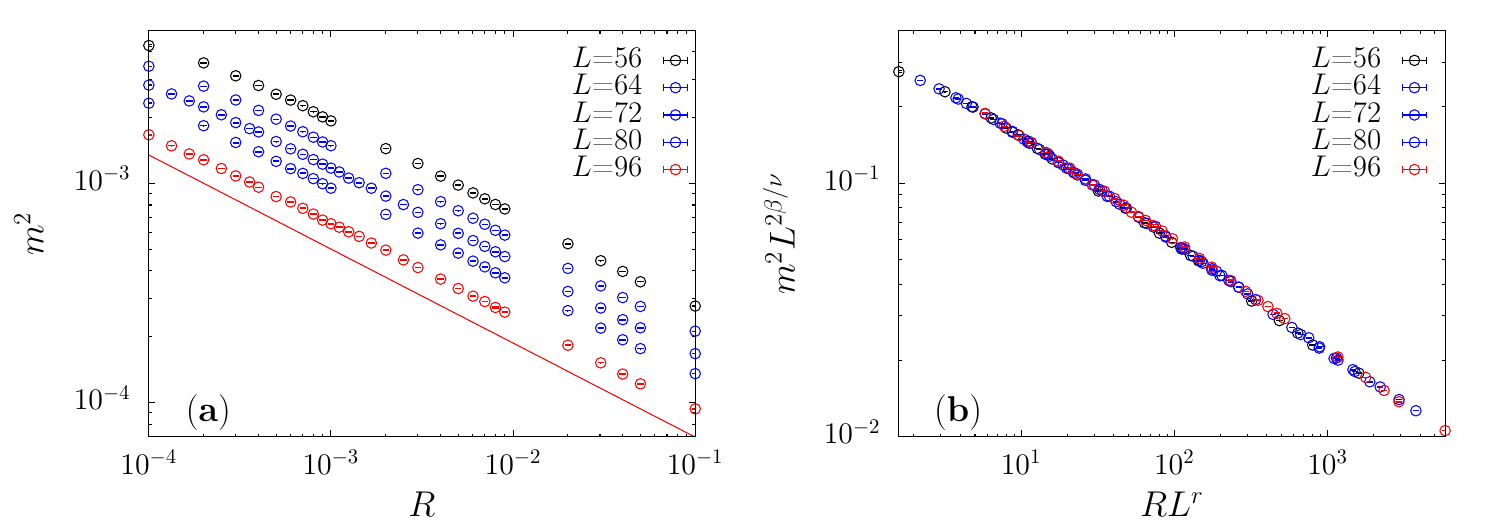}
\caption{ Dimerized  Heisenberg model: driven dynamics of $m^2 $ versus $R$ for different system size $L$ on log-log plots.  From the dimer phase to the QCP using linear   protocol before (a) and after rescaling (b). The red line segments in (a) have slopes of ($2\beta-d\nu$)/($\nu r$). }
\label{fig:j1j2drivengap}
\end{figure}

\begin{figure}[htp]
\centering
\includegraphics[width=0.8\textwidth]{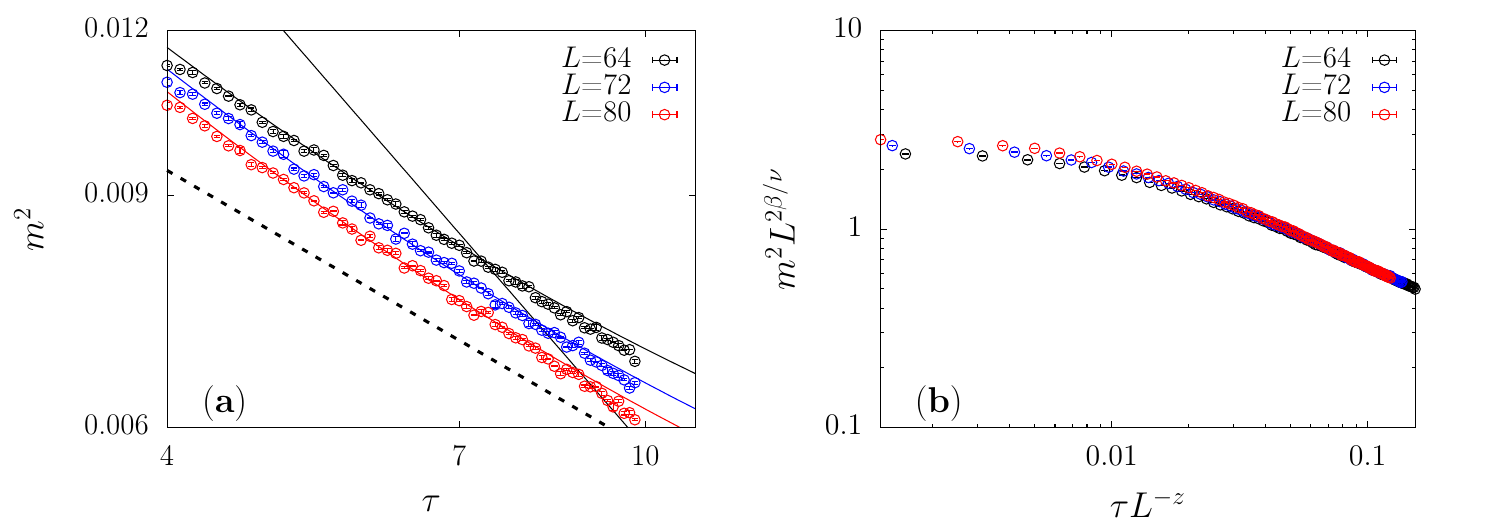}
\caption{Relaxation dynamics of $m^2 $ versus $\tau$ for different system size $L$ on log-log plots. For a sudden quench from the gapless AFM phase to the QCP before (a) and after rescaling (b), the black solid line in (a) follows the scaling relation $ m^2 \propto \tau^{2\beta/(z\nu)} $, while the black dashed lines is given by the modified scaling $ m^2 \propto \tau^{2\beta/(z_{02}\nu)} $.  The red, blue and black curves are fitting lines  obtained using our newly proposed  scaling relation.}
\label{fig:j1j2relaxgapless}
\end{figure}

\subsection{Imaginary time relaxation dynamics}

We then consider the relaxation dynamics. In the relaxation protocol, we prepare the initial state at $J_0=1.90951$ in gapless AFM phase and then perform a sudden quench to the critical point $J=1.0$, studying the subsequent relaxation process.  We find that the  scaling relation $ m^2 \propto \tau^{-2\beta/(z\nu)} $~\cite{Yin2014}, which accounts only for contributions from the critical point with $z=1$, fails to describe the data (as shown by the black solid line in Fig.~\ref{fig:j1j2relaxgapless}(a)). 

Analogous to the driven case, replacing the dynamic critical exponent $ z $ with an effective value of $z_{\text{02}}=2$ yields a reasonable fit to the data (black dashed line in Fig.~\ref{fig:j1j2relaxgapless}(a)), suggesting an apparent deviation from standard scaling. However, this contradicts the data collapse analysis, which clearly shows that setting $ z = 1 $—the known dynamic exponent at the critical point—leads to excellent scaling collapse across different system sizes ( Fig.~\ref{fig:j1j2relaxgapless}(b)). 

\begin{ruledtabular}
\begin{table}[!h]
\caption{Fitting results for the data in Fig.~\ref{fig:j1j2relaxgapless}~(a)  with $ m^2 = q_1\tau^{-2\beta/\nu z} + q_2L^{-1}\tau^{-2\beta/\nu z+1/z} $.	We fixed  known exponents $\beta$, $\nu$ and $z$ in the fit and treated $ q_1 $ and $ q_2 $ as free parameters. }
\begin{tabular}{l c c  }
 	   	   $L$ 	  & $q_1$ 	  & $q_2$ 	 \\
 	   	   	
\hline
	$64$			& 0.032(2)   	&0.27(1)       \\   
	$72$			& 0.032(1)  	&0.28(1)         \\
    $80$			& 0.031(1)   	&0.29(2)       \\
\end{tabular}
\label{sext2}
\end{table}
\end{ruledtabular}

To resolve this discrepancy, and in analogy with our analysis of driven dynamics, we propose a modified single-point relaxation scaling relation that incorporates the influence of the gapless and ordered initial state: $ m^2 = q_1\tau^{-2\beta/\nu z} + q_2L^{-1}\tau^{-2\beta/\nu z+1/z} $. We find that our proposed scaling relation describes the data very well (see Fig.~\ref{fig:j1j2relaxgapless}(a)). Moreover, the fitted parameters $ q_1 $ and $ q_2 $  remain stable within error bars across different system sizes (see Table~\ref{sext2}), providing strong support for the validity of our scaling ansatz.

For comparison, we also consider a sudden quench from the gapped dimer phase ($ J_0 = 0.0 $) to the critical point ($ J = 1.0 $). In this case, the data are well described by the  scaling  relation $ m^2 \propto \tau^{(d - 2\beta/\nu)/z} $ (red solid line in Fig.~\ref{fig:j1j2relaxgap}(a)), and data from all system sizes collapse perfectly when using $ z = 1 $ (Fig.~\ref{fig:j1j2relaxgap}(b)).

\begin{figure}[htp]
\centering
\includegraphics[width=0.8\textwidth]{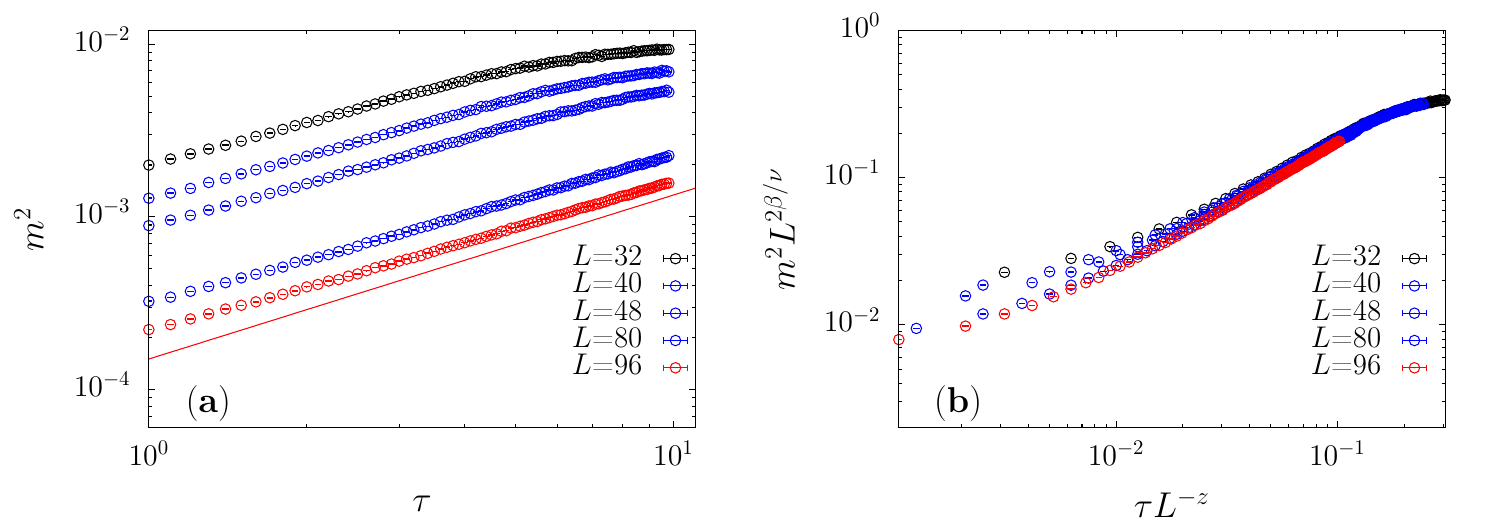}
\caption{Relaxation dynamics of $m^2 $ versus $\tau$ for different system size $L$ on log-log plots.  For a sudden quench from the gapped dimer initial state to the  QCP before (a) and after rescaling (b), the red line in (a) is described by the  scaling relation $ m^2 \propto \tau^{(d - 2\beta/\nu)/z} $. }
\label{fig:j1j2relaxgap}
\end{figure}

\end{widetext}

\end{document}